\documentclass[preprint, superscriptaddress,preprintnumbers,amsmath,amssymb,pra]{revtex4}
\usepackage{graphicx}

\begin{document}
\newcommand{\sil}{\sigma_{\|}}
\newcommand{\sit}{\sigma_{\bot}}
\newcommand{\sila}{\sigma_{\|}(\omega,k,T,\Delta,\mu)}
\newcommand{\sita}{\sigma_{\bot}(\omega,k,T,\Delta,\mu)}
\newcommand{\ar}{(\omega,k,T,\Delta,\mu)}
\newcommand{\ho}{\hbar\omega}

\thispagestyle{empty}

\title{Conductivity of graphene in the framework of Dirac model:
Interplay between nonzero mass gap and chemical potential}

\author{
G.~L.~Klimchitskaya}
\affiliation{Central Astronomical Observatory at Pulkovo of the
Russian Academy of Sciences, Saint Petersburg,
196140, Russia}
\affiliation{Institute of Physics, Nanotechnology and
Telecommunications, Peter the Great Saint Petersburg
Polytechnic University, Saint Petersburg, 195251, Russia}

\author{
V.~M.~Mostepanenko}
\affiliation{Central Astronomical Observatory at Pulkovo of the
Russian Academy of Sciences, Saint Petersburg,
196140, Russia}
\affiliation{Institute of Physics, Nanotechnology and
Telecommunications, Peter the Great Saint Petersburg
Polytechnic University, Saint Petersburg, 195251, Russia}
\affiliation{Kazan Federal University, Kazan, 420008, Russia}

\author{
V.~M.~Petrov}
\affiliation{Institute of Physics, Nanotechnology and
Telecommunications, Peter the Great Saint Petersburg
Polytechnic University, Saint Petersburg, 195251, Russia}

\begin{abstract}
The complete theory of electrical conductivity of graphene at arbitrary
temperature is developed with taken into account mass-gap parameter and
chemical potential. Both the in-plane and out-of-plane conductivities
of graphene are expressed via the components of the polarization tensor
in (2+1)-dimensional space-time analytically continued to the real
frequency axis. Simple analytic expressions for both the real and
imaginary parts of the conductivity of graphene are obtained at zero
and nonzero temperature. They demonstrate an interesting interplay
depending on the values of mass gap and chemical potential. In the
local limit, several results obtained earlier using various approximate
and phenomenological approaches are reproduced, refined and generalized.
The numerical computations of both the real and imaginary parts of the
conductivity of graphene are performed to illustrate the obtained results.
The analytic expressions for the conductivity of graphene obtained in this
paper can serve as a guide in the comparison between different theoretical
approaches and between experiment and theory.
\end{abstract}

\maketitle

\section{INTRODUCTION}

The electrical conductivity of graphene, a two-dimensional hexagonal lattice
of carbon atoms, possesses many qualities required for prospective applications
in both fundamental physics and nanotechnology \cite{1,2}.
At energies below a few electron volts the electronic properties of graphene
are well described by the Dirac model \cite{1,2,3}.
In the context of this model graphene quasiparticles obey a linear dispersion
relation, where the speed of light $c$ is replaced with the Fermi velocity
$v_F\approx c/300$. The electrical properties of graphene are essentially connected
with an existence of the so-called universal conductivity $\sigma_0$ expressed
via the electron charge $e$ and Planck constant $\hbar$. For some time, a few
distinct values for $\sigma_0$ have been obtained by different authors \cite{4,5,6,7}.
Eventually a consensus was reached on the expression $\sigma_0=e^2/(4\hbar)$
\cite{7,8,9,10,11,12,13}.

In general, the conductivity of graphene is nonlocal, i.e., depends on both the
frequency and the magnitude of the wave vector, and also on the temperature.
It was investigated by many authors using the phenomenological two-dimensional
Drude model, the current-current correlation function in the random phase
approximation, the Kubo response formalism, and the Boltzmann transport equation
\cite{1,2,3,4,5,6,7,9,10,11,12,13,14,15,16,17,18,19,20,21,22,23,24,25,26,27,27a,28,29,30,31,32,33,34,35}.
 Furthermore, the electrical conductivity of graphene is sensitive to whether the mass
of quasiparticles $m$ is exactly equal to zero or it is rather small but nonzero.
For real graphene samples a nonzero mass gap $\Delta=2mc^2$ in the energy spectrum
of quasiparticles arises under the influence of electron-electron interactions,
substrates and impurities \cite{3,26,36,37}
Some partial results for the conductivity of gapped graphene have been obtained
using the two-band model \cite{24} and the static polarization function \cite{34}.
In the local approximation at zero temperature the conductivity of gapped graphene was
also examined in Refs.~\cite{12,21,22}.

Real graphene samples are always doped and can be characterized by some nonzero
chemical potential $\mu$. The electrical conductivity of doped graphene was explored
by using several approximate methods \cite{12,14,24,25,29,34}.
The question arises of whether there are significant differences in impacts of
the nonzero mass-gap parameter and chemical potential on the conductivity of graphene.
On this subject it was noticed \cite{12} that at zero temperature in the local
approximation the response function for undoped but gapped graphene is similar to
the case of doped but ungapped graphene if to identify the gap parameter,
$\Delta$, with twice the Fermi energy, $2E_F$.

A comprehensive investigation of the conductivity of graphene can be performed
using the exact expression for its polarization tensor at any temperature, mass gap
and chemical potential. Although in some specific cases the polarization tensor in
(2+1)-dimensional space-time has been calculated by many authors (see, e.g.,
Refs.~\cite{14,26,36,38}), the complete results needed for a fundamental understanding
of the conductivity were obtained only recently.
In Ref.~\cite{39} the exact polarization tensor of graphene with any mass-gap
parameter has been found at zero temperature. The extension of this tensor to the
case of nonzero temperature was made in Ref.~\cite{40}, but only at the pure
imaginary Matsubara frequencies. The results of Refs.~\cite{39,40} have been
extensively used to calculate the Casimir force in graphene systems
\cite{41,42,43,44,45,46,47,48,49,50,51},
but they are not directly applicable in the studies of conductivity which is defined
along the real frequency axis.

Another representation for the polarization tensor of gapped graphene, allowing
an analytic continuation to the real frequency axis, was derived in Ref.~\cite{52}.
It was applied in calculations of the Casimir force \cite{53,54,55,56}, on the one
hand, and of the reflectivity properties of graphene and graphene-coated substrates
\cite{52,57,58,59}, on the other hand. For the latter purposes, explicit expressions
for the polarization tensor at real frequencies have been obtained for a gapless
\cite{52,57} and gapped \cite{58} graphene. This has opened up opportunities for
a detailed study of the conductivity of graphene on the basis of first principles
of quantum electrodynamics at nonzero temperature. The conductivity of pure
(gapless)   graphene was investigated in Ref.~\cite{60} using the continuation
of the polarization tensor to real frequencies derived in Refs.~\cite{52,57}.
The case of gapped graphene was considered in Ref.~\cite{61} with the help of
analytic continuation of the polarization tensor obtained for this case in Ref.~\cite{58}.
In so doing, the previously known partial results for the conductivity of graphene
have been reproduced and their generalizations to the case of any temperature with
taken into account effects of nonlocality have been obtained.

In this paper, we develop the complete theory for the electrical conductivity
of graphene in the framework of the Dirac model at arbitrary values of the mass gap,
temperature and chemical potential. For this purpose, the results of Ref.~\cite{62}
are used, where the polarization tensor of graphene of Ref.~\cite{52} was
generalized to the case of graphene with nonzero chemical potential.
We perform an analytic continuation of the polarization tensor of Ref.~\cite{62}
to the real frequency axis and express both the longitudinal (in-plane) and
transverse (perpendicular to the plane of graphene) conductivities in terms of its
component. Then, the impact of chemical potential on the real part of conductivity
at both zero and nonzero temperature is investigated. It is shown that at low
frequencies satisfying a condition $\hbar\omega<\Delta$ the real part of conductivity
of graphene is equal to zero at any values of temperature and chemical potential.
For $\hbar\omega\geq\Delta$ the real part of conductivity
is not equal to zero and there is an interesting interplay
depending on the specific values of $\Delta$ and $\mu$.
Thus, at zero temperature the real part of conductivity vanishes for $\hbar\omega<2\mu$
and is a simple function of $\sigma_0$, $\omega$, and $\Delta$ for $\hbar\omega> 2\mu$.
At nonzero temperature the real part of conductivity depends explicitly on
$\sigma_0$, $\omega$, $T$, $\Delta$, and $\mu$. For the imaginary part of conductivity,
simple analytic expressions are obtained at zero temperature, and the results of
numerical computations at nonzero temperature are presented again demonstrating an
important interplay between the values of $\Delta$ and $\mu$. In all cases we consider
both the exact results and their local limits. The obtained analytic expressions are
compared with those found earlier in the literature in the framework of various
theoretical approaches.

The paper is organized as follows. In Sec.~II we present an exact formalism
expressing the conductivity of gapped graphene with nonzero chemical potential
at any temperature via the components of the polarization tensor defined along
the real frequency axis. Section~III investigates the impact of chemical potential
on the real part of conductivity. Here, the imaginary part of the polarization
tensor is expanded in powers of a small parameter. Both cases of nonzero and
zero temperature are considered. In Sec.~IV  the impact of chemical potential
on the imaginary part of conductivity of graphene is examined. Here, the main
contribution to the real part of the polarization tensor is found  and the
cases of nonzero and zero temperature are considered.
In Sec.~V the reader will find our conclusions and a discussion of the obtained
results. The Appendix contains the proof of several asymptotic relations used
in Sec.~IV.

\section{Exact formalism for arbitrary temperature, mass-gap parameter
and chemical potential}

The polarization tensor allows complete description for the response of a physical
system  to external electromagnetic field. Although the tensor used below is found
in the one-loop approximation, we call the formalism under consideration {\it exact}
implying that in the framework of this approximation the role of nonzero temperature,
mass gap and chemical potential is taken into account precisely.
We denote the polarization tensor of graphene as
$\Pi_{mn}(\omega,k,T,\Delta,\mu)$, where $m,\,n=0,\,1,\,2$,
$k=|\mbox{\boldmath$k$}|$ is the magnitude of the wave vector component parallel
to graphene, and $T$ is the temperature. The standard definition of $\Pi_{mn}$
in terms of the spinor propagator can be found in the textbooks \cite{63,64}
and, in application to graphene, in Refs.~\cite{39,40,60}.

It is known \cite{39,40} that only the two components of the polarization tensor
are the independent quantities. In the literature the most frequently used
quantities are $\Pi_{00}$ and $\Pi_{\rm tr}\equiv\Pi_{n}^{\,n}$.
Instead of the latter, however, for our purposes it is more convenient to employ
the following quantity:
\begin{equation}
\Pi(\omega,k,T,\Delta,\mu)=k^2\Pi_{\rm tr}(\omega,k,T,\Delta,\mu)+
\left(\frac{\omega^2}{c^2}-k^2\right)\Pi_{00}(\omega,k,T,\Delta,\mu).
\label{eq1}
\end{equation}

A knowledge of the polarization tensor allows one to find the conductivity of
graphene. The reason is that both the longitudinal, $\sil$, and transverse, $\sit$,
conductivities of graphene can be expressed via the respective correlation
functions (see, e.g., Refs.~\cite{65,66}). The latter in turn are directly connected
with the polarization tensor \cite{50}. Eventually, the in-plane and out-of-plane
conductivities of graphene are given by \cite{50,60,61}
\begin{eqnarray}
&&
\sila=-i\frac{\omega}{4\pi\hbar k^2}\Pi_{00}{\ar},
\nonumber \\
&&
\sita=i\frac{c^2}{4\pi\hbar\omega k^2}\Pi{\ar}.
\label{eq2}
\end{eqnarray}

To find the conductivities of graphene from Eq.~({\ref{eq2}), we continue the
polarization tensor of graphene with nonzero $\Delta$ and $\mu$ found in
Ref.~\cite{62} to the real frequency axis. Following Ref.~\cite{62}, it is
convenient to present the quantities $\Pi_{00}$ and $\Pi$ as the sums of two
contributions
\begin{eqnarray}
&&
\Pi_{00}(\omega,k,T,\Delta,\mu)=\Pi_{00}^{(0)}(\omega,k,\Delta)+
\Pi_{00}^{(1)}(\omega,k,T,\Delta,\mu),
\nonumber \\
&&
\Pi(\omega,k,T,\Delta,\mu)=\Pi^{(0)}(\omega,k,\Delta)+
\Pi^{(1)}(\omega,k,T,\Delta,\mu).
\label{eq3}
\end{eqnarray}
\noindent
Here, the terms $\Pi_{00}^{(0)}$ and $\Pi^{(0)}$ describe the gapped
($\Delta\neq 0$) but undoped ($\mu=0$) graphene at $T=0$.
The explicit expressions for these terms were obtained in Ref.~\cite{39}
(see also Refs.~\cite{52,61} for an explicit analytic continuation to the
real frequency axis). They can be written in the form
\begin{eqnarray}
&&
\Pi_{00}^{(0)}(\omega,k,\Delta)=-\frac{2\alpha k^2c}{\omega^2\eta^2}
\,\tilde{\Phi}(\omega,k,\Delta),
\nonumber \\
&&
\Pi^{(0)}(\omega,k,\Delta)=\frac{2\alpha k^2}{c}
\,\tilde{\Phi}(\omega,k,\Delta),
\label{eq4}
\end{eqnarray}
\noindent
where $\alpha=e^2/(\hbar c)$ is the fine structure constant and the quantity
$\eta$ is defined as
\begin{equation}
\eta=\eta(\omega,k)=\sqrt{1-\frac{v_F^2k^2}{\omega^2}}.
\label{eq5}
\end{equation}
\noindent
The function $\tilde{\Phi}$ along the axis of real frequencies is given by
\cite{52,61}
\begin{equation}
\tilde{\Phi}(\omega,k,\Delta)=\left\{
\begin{array}{ll}
\Delta-\hbar\omega\eta\left[1+\left(\frac{\Delta}{\hbar\omega\eta}\right)^2\right]
{\rm arctanh}\frac{\hbar\omega\eta}{\Delta}, & \hbar\omega\eta<\Delta,\\[3mm]
\Delta-\hbar\omega\eta\left[1+\left(\frac{\Delta}{\hbar\omega\eta}\right)^2\right]
\left({\rm arctanh}\frac{\Delta}{\hbar\omega\eta}
+i\frac{\pi}{2}\right), & \hbar\omega\eta\geq\Delta.
\end{array}\right.
\label{eq6}
\end{equation}
\noindent
Note that this $\tilde{\Phi}$ is different from the function $\Phi$ in
Refs.~\cite{52,61} by a factor, which is taken into account in Eq.~(\ref{eq4}).
According to  Eqs.~(\ref{eq4}) and (\ref{eq6}), the quantities $\Pi_{00}^{(0)}$
and $\Pi^{(0)}$ are real if $\hbar\omega\eta<\Delta$ and have both the real and
imaginary parts if $\hbar\omega\eta\geq\Delta$.

The terms $\Pi_{00}^{(1)}$ and $\Pi^{(1)}$ on the right-hand sides of Eq.~(\ref{eq3})
take into consideration the dependences on the temperature and on the chemical
potential. Note that these terms should not be confused with the thermal corrections
because they may remain not equal to zero in the limiting case $T\to 0$
(see Secs.~III and IV). The resulting contributions, which may also depend on
$\Delta$, describe a dependence of the zero-temperature polarization tensor on
the chemical potential. They should vanish in the limiting case $\mu\to 0$.

The explicit expressions for  $\Pi_{00}^{(1)}$ and $\Pi^{(1)}$,
found in Ref.~\cite{62},
were analytically continued to the real frequency axis following Refs.~\cite{52,58}.
For the imaginary parts of $\Pi_{00}^{(1)}$ and $\Pi^{(1)}$, the results are
\begin{eqnarray}
&&
{\rm Im}\Pi_{00}^{(1)}{\ar}=\frac{4\alpha\hbar c^2}{v_F^2\omega\eta}\,
\theta(\ho\eta-\Delta)
\nonumber \\
&&~~
\times\int_{u^{(-)}}^{u^{(+)}}\!\!du\left(\frac{1}{e^{\beta u+\frac{\mu}{k_BT}}+1}+
\frac{1}{e^{\beta u-\frac{\mu}{k_BT}}+1}\right)
\frac{(2cu-\omega)^2-v_F^2k^2}{\sqrt{v_F^2k^2A(\omega,k)-(2cu-\omega)^2}},
\nonumber \\
&&
{\rm Im}\Pi^{(1)}{\ar}=\frac{4\alpha\hbar\omega\eta}{v_F^2}\,
\theta(\ho\eta-\Delta)
\label{eq7} \\
&&~~
\times\int_{u^{(-)}}^{u^{(+)}}\!\!du\left(\frac{1}{e^{\beta u+\frac{\mu}{k_BT}}+1}+
\frac{1}{e^{\beta u-\frac{\mu}{k_BT}}+1}\right)
\frac{(2cu-\omega)^2+v_F^2k^2[1-A(\omega,k)]}{\sqrt{v_F^2k^2A(\omega,k)-(2cu-\omega)^2}}.
\nonumber
\end{eqnarray}
\noindent
Here, $\beta=\hbar c/(k_BT)$, $\theta(x)$ is the step function equal to unity for
$x\geq 0$ and zero for $x<0$, and the following notations are introduced
\begin{eqnarray}
&&
u^{(\pm)}=\frac{1}{2c}[\omega\pm v_Fk\sqrt{A(\omega,k)}],
\nonumber \\
&&A(\omega,k)=1-\left(\frac{\Delta}{\ho\eta}\right)^2.
\label{eq8}
\end{eqnarray}
\noindent
Note that not only $u^{(+)}>0$ but, due to the condition $ck<\omega$, it is valid that
$u^{(-)}>0$ as well.

The real parts of $\Pi_{00}^{(1)}$ and $\Pi^{(1)}$ on the right-hand sides of
Eq.~(\ref{eq3}) are defined differently in different frequency regions.
For the sake of brevity, we introduce the notations
\begin{eqnarray}
&&
B_1(x)=\frac{x^2-v_F^2k^2}{\sqrt{x^2-v_F^2k^2A(\omega,k)}},
\label{eq9} \\
&&
B_2(x)=\frac{x^2+v_F^2k^2[1-A(\omega,k)]}{\sqrt{x^2-v_F^2k^2A(\omega,k)}}.
\nonumber
\end{eqnarray}
\noindent
Then, for the frequencies satisfying the condition $\ho\eta<\Delta$, one obtains
\begin{eqnarray}
&&
{\rm Re}\Pi_{00}^{(1)}{\ar}=\frac{8\alpha\hbar c^2}{v_F^2}
\int_{\Delta/(2\hbar c)}^{\infty}\!\!\!du\sum_{\kappa=\pm 1}
\frac{1}{e^{\beta u+\kappa\frac{\mu}{k_BT}}+1}
\nonumber \\
&&~~
\times\left\{1-\frac{1}{2\omega\eta}\left[B_1(2cu+\omega)-
B_1(2cu-\omega)\right]\right\},
\nonumber \\
&&
{\rm Re}\Pi^{(1)}\ar=\frac{8\alpha\hbar \omega^2}{v_F^2}
\int_{\Delta/(2\hbar c)}^{\infty}\!\!\!du\sum_{\kappa=\pm 1}
\frac{1}{e^{\beta u+\kappa\frac{\mu}{k_BT}}+1}
\nonumber \\
&&~~
\times\left\{1-\frac{\eta}{2\omega}\left[B_2(2cu+\omega)-
B_2(2cu-\omega)\right]\right\}.
\label{eq10}
\end{eqnarray}
\noindent
Here and below we use an abbreviated notation for the sum of exponent-containing
fractions which is written out in full in Eq.~(\ref{eq7}).

Now we present the real parts of $\Pi_{00}^{(1)}$ and $\Pi^{(1)}$ in the frequency
region $\ho\eta\geq\Delta$. Here, the convenient expression for
${\rm Re}\Pi_{00}^{(1)}$ is the following:
\begin{equation}
{\rm Re}\Pi_{00}^{(1)}\ar=\frac{8\alpha\hbar c^2}{v_F^2}(I_1+I_2+I_3),
\label{eq11}
\end{equation}
\noindent
where the integrals $I_n$ are given by
\begin{eqnarray}
&&
I_1=\int_{\Delta/(2\hbar c)}^{u^{(-)}}\!\!\!du\sum_{\kappa=\pm 1}
\frac{1}{e^{\beta u+\kappa\frac{\mu}{k_BT}}+1}
\nonumber \\
&&~~
\times\left\{1-\frac{1}{2\omega\eta}\left[B_1(2cu+\omega)+
B_1(2cu-\omega)\right]\right\},
\nonumber\\
&&
I_2=\int_{u^{(-)}}^{u^{(+)}}\!\!\!du\sum_{\kappa=\pm 1}
\frac{1}{e^{\beta u+\kappa\frac{\mu}{k_BT}}+1}
\nonumber \\
&&~~
\times\left[1-\frac{1}{2\omega\eta}B_1(2cu+\omega)\right],
\label{eq12} \\
&&
I_3=\int_{u^{(+)}}^{\infty}\!\!\!du\sum_{\kappa=\pm 1}
\frac{1}{e^{\beta u+\kappa\frac{\mu}{k_BT}}+1}
\nonumber \\
&&~~
\times\left\{1-\frac{1}{2\omega\eta}\left[B_1(2cu+\omega)-
B_1(2cu-\omega)\right]\right\}.
\nonumber
\end{eqnarray}

A similar expression for ${\rm Re}\Pi^{(1)}$ in the interval  $\ho\eta\geq\Delta$
reads
\begin{equation}
{\rm Re}\Pi^{(1)}\ar=\frac{8\alpha\hbar \omega^2}{v_F^2}(J_1+J_2+J_3),
\label{eq13}
\end{equation}
\noindent
where
\begin{eqnarray}
&&
J_1=\int_{\Delta/(2\hbar c)}^{u^{(-)}}\!\!\!du\sum_{\kappa=\pm 1}
\frac{1}{e^{\beta u+\kappa\frac{\mu}{k_BT}}+1}
\nonumber \\
&&~~
\times\left\{1-\frac{\eta}{2\omega}\left[B_2(2cu+\omega)+
B_2(2cu-\omega)\right]\right\},
\nonumber\\
&&
J_2=\int_{u^{(-)}}^{u^{(+)}}\!\!\!du\sum_{\kappa=\pm 1}
\frac{1}{e^{\beta u+\kappa\frac{\mu}{k_BT}}+1}
\nonumber \\
&&~~
\times\left[1-\frac{\eta}{2\omega}B_2(2cu+\omega)\right],
\label{eq14} \\
&&
J_3=\int_{u^{(+)}}^{\infty}\!\!\!du\sum_{\kappa=\pm 1}
\frac{1}{e^{\beta u+\kappa\frac{\mu}{k_BT}}+1}
\nonumber \\
&&~~
\times\left\{1-\frac{\eta}{2\omega}\left[B_2(2cu+\omega)-
B_2(2cu-\omega)\right]\right\}.
\nonumber
\end{eqnarray}

Note that Eqs.~(\ref{eq7}), (\ref{eq10}), (\ref{eq12}), and (\ref{eq14}), which
take into account  the chemical potential of gapped graphene, are also immediately
obtainable from the respective equations of Ref.~\cite{61}, devoted to the conductivity
of gapped but undoped graphene, by the substitution
\begin{equation}
\frac{1}{e^{\beta u}+1}\to\frac{1}{2}
\sum_{\kappa=\pm 1}
\frac{1}{e^{\beta u+\kappa\frac{\mu}{k_BT}}+1}.
\label{eq15}
\end{equation}
\noindent
This substitution is in fact the standard procedure for introducing the chemical
potential in thermal quantum field theory \cite{67}.

As a result, graphene conductivities are presented by Eq.~(\ref{eq2}), where
the quantities $\Pi_{00}$ and $\Pi$ are given by
Eqs.~(\ref{eq3}), (\ref{eq4}), (\ref{eq7}), (\ref{eq10}), (\ref{eq11}),
and (\ref{eq13}). These equations are used below to investigate an interplay
between the mass gap and chemical potential and their combine impact on the
conductivity properties.

\section{Impact of chemical potential on real part of conducti\-vity}

According to the results of Sec.~II, the polarization tensor of graphene along
the real frequency axis is a complex function, i.e., it has both the real and
imaginary parts. {}From Eq.~(\ref{eq2}) for the real parts of conductivities one
obtains
\begin{eqnarray}
&&
{\rm Re}\sila=\frac{\omega}{4\pi\hbar k^2}{\rm Im}\Pi_{00}{\ar},
\nonumber \\
&&
{\rm Re}\sita=-\frac{c^2}{4\pi\hbar\omega k^2}{\rm Im}\Pi{\ar}.
\label{eq16}
\end{eqnarray}
\noindent
In agreement with Eq.~({\ref{eq3})
\begin{eqnarray}
&&
{\rm Im}\Pi_{00}(\omega,k,T,\Delta,\mu)=
{\rm Im}\Pi_{00}^{(0)}(\omega,k,\Delta)+
{\rm Im}\Pi_{00}^{(1)}(\omega,k,T,\Delta,\mu),
\nonumber \\
&&
{\rm Im}\Pi(\omega,k,T,\Delta,\mu)={\rm Im}\Pi^{(0)}(\omega,k,\Delta)+
{\rm Im}\Pi^{(1)}(\omega,k,T,\Delta,\mu).
\label{eq17}
\end{eqnarray}
\noindent
Here, the first contributions on the right-hand sides are found from
Eqs.~(\ref{eq4}) and  (\ref{eq6})
\begin{eqnarray}
&&
{\rm Im}\Pi_{00}^{(0)}(\omega,k,\Delta)=
\frac{\pi\alpha\hbar ck^2}{\omega\eta}\theta(\ho\eta-\Delta)
\left[1+\left(\frac{\Delta}{\ho\eta}\right)^2\right],
\label{eq18} \\[2mm]
&&
{\rm Im}\Pi^{(0)}(\omega,k,\Delta)=
-\frac{\pi\alpha\hbar\omega k^2\eta}{c}\theta(\ho\eta-\Delta)
\left[1+\left(\frac{\Delta}{\ho\eta}\right)^2\right].
\nonumber
\end{eqnarray}

Using  Eq.~(\ref{eq5}) we expand Eq.~(\ref{eq18}) in powers of a small parameter
$v_F/c\sim 3\times 10^{-3}$ preserving only the terms up to the second order
\begin{eqnarray}
&&
{\rm Im}\Pi_{00}^{(0)}(\omega,k,\Delta)=
\frac{\pi\alpha\hbar ck^2}{\omega}\theta(\ho\eta-\Delta)
\left[\frac{(\ho)^2+\Delta^2}{(\ho)^2}+\frac{v_F^2k^2}{2\omega^2}
\frac{(\ho)^2+3\Delta^2}{(\ho)^2}\right],
\label{eq19} \\[2mm]
&&
{\rm Im}\Pi^{(0)}(\omega,k,\Delta)=
-\frac{\pi\alpha\hbar\omega k^2}{c}\theta(\ho\eta-\Delta)
\left[\frac{(\ho)^2+\Delta^2}{(\ho)^2}-\frac{v_F^2k^2}{2\omega^2}
\frac{(\ho)^2-\Delta^2}{(\ho)^2}\right].
\nonumber
\end{eqnarray}
\noindent
Taking into account that
\begin{equation}
\frac{v_F^2k^2}{\omega^2}=\frac{v_F^2}{c^2}\frac{c^2k^2}{\omega^2}
<\frac{v_F^2}{c^2},
\label{eq20}
\end{equation}
\noindent
the nonlocal contributions in Eq.~(\ref{eq19}) are much smaller, as compared
to the local ones.

The second contributions to Eq.~(\ref{eq17}) are given by Eq.~(\ref{eq7}).
As can be seen from Eqs.~(\ref{eq7}) and (\ref{eq18}), the imaginary parts
of $\Pi_{00}$ and $\Pi$ and, thus, the real parts of the conductivity of
graphene (\ref{eq16}) are not equal to zero only in the frequency region
satisfying a condition $\ho\eta\geq\Delta$. Below we obtain the expansions
of ${\rm Im}\Pi_{00}^{(1)}$ and ${\rm Im}\Pi^{(1)}$ in powers of $v_F^2/c^2$
similar to  Eq.~(\ref{eq19}).

\subsection{Imaginary parts of contributions to the polarization tensor
depending on chemical potential}

To obtain the desired expansions, we start from Eq.~(\ref{eq7}) and introduce
the new integration variable
\begin{equation}\tau=\frac{2cu-\omega}{v_Fk\sqrt{A(\omega,k)}}.
\label{eq21}
\end{equation}
\noindent
Then Eq.~(\ref{eq7}) takes the form
\begin{eqnarray}
&&
{\rm Im}\Pi_{00}^{(1)}{\ar}=-\frac{2\alpha\hbar ck^2}{\omega\eta}\,
\theta(\ho\eta-\Delta)\!\int_{-1}^{1}\!\!\!d\tau
\sum_{\kappa=\pm 1}\frac{1}{e^{\frac{\ho+2\kappa\mu}{2k_BT}+\gamma\tau}+1}
\frac{1-A(\omega,k)\tau^2}{\sqrt{1-\tau^2}},
\label{eq22} \\
&&
{\rm Im}\Pi^{(1)}{\ar}=\frac{2\alpha\hbar\omega\eta k^2}{c}\,
\theta(\ho\eta-\Delta)\!\int_{-1}^{1}\!\!\!d\tau
\sum_{\kappa=\pm 1}\frac{1}{e^{\frac{\ho+2\kappa\mu}{2k_BT}+\gamma\tau}+1}
\frac{1-A(\omega,k)(1-\tau^2)}{\sqrt{1-\tau^2}}
.
\nonumber
\end{eqnarray}
\noindent
Here, the following notation is introduced:
\begin{equation}
\gamma=\gamma(\omega,k,\Delta)=\frac{v_Fk}{\omega}
\frac{\ho}{2k_BT}\sqrt{A(\omega,k)}.
\label{eq23}
\end{equation}

Now we preserve only the terms up to the second order in $v_F/c$ so that
\begin{eqnarray}
&&
\eta=1-\frac{v_F^2k^2}{2\omega^2},\quad
\frac{1}{\eta}=1+\frac{v_F^2k^2}{2\omega^2},
\label{eq24} \\
&&
A(\omega,k)=A_0- \frac{v_F^2k^2\Delta^2}{\hbar^2\omega^4}, \quad
A_0\equiv 1-\left(\frac{\Delta}{\ho}\right)^2.
\nonumber
\end{eqnarray}
\noindent
Expanding also the exponent-containing fractions up to the second order in
the small parameter $\gamma\tau\sim v_F/c$, one can rewrite Eq.~(\ref{eq22})
in the form
\begin{eqnarray}
&&
{\rm Im}\Pi_{00}^{(1)}{\ar}=-\frac{2\alpha\hbar ck^2}{\omega}\,
\theta(\ho\eta-\Delta)
\sum_{\kappa=\pm 1}\left\{
\frac{1}{e^{\frac{\ho+2\kappa\mu}{2k_BT}}+1}\int_{-1}^{1}\!\!\!d\tau
\frac{1-A_0\tau^2}{\sqrt{1-\tau^2}}\right.
\nonumber \\
&&~~~
+\frac{v_F^2k^2}{2\omega^2}\left[
\frac{1}{e^{\frac{\ho+2\kappa\mu}{2k_BT}}+1}\int_{-1}^{1}\!\!\!d\tau
\frac{1-A_0\tau^2}{\sqrt{1-\tau^2}}
+2\left(\frac{\Delta}{\ho}\right)^2
\frac{1}{e^{\frac{\ho+2\kappa\mu}{2k_BT}}+1}\int_{-1}^{1}\!\!\!d\tau
\frac{\tau^2}{\sqrt{1-\tau^2}}
\right.
\nonumber \\
&&~~~~
+2A_0\left(\frac{\ho}{2k_BT}\right)^2
{\rm csch}^3\frac{\ho+2\kappa\mu}{2k_BT}
{\rm sinh}^4\frac{\ho+2\kappa\mu}{4k_BT}
\left.\left.
\int_{-1}^{1}\!\!\!d\tau\tau^2
\frac{1-A_0\tau^2}{\sqrt{1-\tau^2}}
\vphantom{\frac{1}{e^{\frac{\ho+2\kappa\mu}{2k_BT}}}}
\right]\right\},
\label{eq25}
\\
&&
{\rm Im}\Pi^{(1)}{\ar}=\frac{2\alpha\hbar\omega k^2}{c}\,
\theta(\ho\eta-\Delta)
\sum_{\kappa=\pm 1}\left\{
\frac{1}{e^{\frac{\ho+2\kappa\mu}{2k_BT}}+1}\int_{-1}^{1}\!\!\!d\tau
\frac{1-A_0(1-\tau^2)}{\sqrt{1-\tau^2}}
\right.
\nonumber \\
&&~~
+\frac{v_F^2k^2}{2\omega^2}\left[
-\frac{1}{e^{\frac{\ho+2\kappa\mu}{2k_BT}}+1}\int_{-1}^{1}\!\!\!d\tau
\frac{1-A_0(1-\tau^2)}{\sqrt{1-\tau^2}}
+
2\left(\frac{\Delta}{\ho}\right)^2
\frac{1}{e^{\frac{\ho+2\kappa\mu}{2k_BT}}+1}\int_{-1}^{1}\!\!d\tau
\sqrt{1-\tau^2}\right.
\nonumber \\
&&~~~~
\left.\left.
+2A_0\left(\frac{\ho}{2k_BT}\right)^2
{\rm csch}^3\frac{\ho+2\kappa\mu}{2k_BT}
{\rm sinh}^4\frac{\ho+2\kappa\mu}{4k_BT}
\int_{-1}^{1}\!\!d\tau\tau^2
\frac{1-A_0(1-\tau^2)}{\sqrt{1-\tau^2}}
\vphantom{\frac{1}{e^{\frac{\ho+2\kappa\mu}{2k_BT}}}}
\right]\right\}.
\nonumber
\end{eqnarray}
\noindent
Note that we have omitted the linear in $\gamma\tau$ terms which lead to zero
results after an integration with respect to $\tau$.

Calculating all integrals in Eq.~({\ref{eq25}), one arrives at
\begin{eqnarray}
&&
{\rm Im}\Pi_{00}^{(1)}{\ar}=-\frac{\pi\alpha\hbar ck^2}{\omega}\,
\theta(\ho\eta-\Delta)
\sum_{\kappa=\pm 1}\left\{
\frac{(\ho)^2+\Delta^2}{(\ho)^2}
\frac{1}{e^{\frac{\ho+2\kappa\mu}{2k_BT}}+1}\right.
\nonumber \\
&&~~~
+\frac{v_F^2k^2}{2\omega^2}\frac{(\ho)^2+3\Delta^2}{(\ho)^2}   \left[
\frac{1}{e^{\frac{\ho+2\kappa\mu}{2k_BT}}+1}\right.
+\left.\left.
\frac{(\ho)^2-\Delta^2}{2(2k_BT)^2}
{\rm csch}^3\frac{\ho+2\kappa\mu}{2k_BT}
{\rm sinh}^4\frac{\ho+2\kappa\mu}{4k_BT}
\right]\right\},
\nonumber \\
&&
{\rm Im}\Pi^{(1)}{\ar}=\frac{\pi\alpha\hbar\omega k^2}{c}\,
\theta(\ho\eta-\Delta)
\sum_{\kappa=\pm 1}\left\{
\frac{(\ho)^2+\Delta^2}{(\ho)^2}
\frac{1}{e^{\frac{\ho+2\kappa\mu}{2k_BT}}+1}\right.
\label{eq26} \\
&&~~~
-\frac{v_F^2k^2}{2\omega^2}\frac{(\ho)^2-\Delta^2}{(\ho)^2}   \left[
\frac{1}{e^{\frac{\ho+2\kappa\mu}{2k_BT}}+1}\right.
-\left.\left.
\frac{3(\ho)^2+\Delta^2}{2(2k_BT)^2}
{\rm csch}^3\frac{\ho+2\kappa\mu}{2k_BT}
{\rm sinh}^4\frac{\ho+2\kappa\mu}{4k_BT}
\right]\right\}.
\nonumber
\end{eqnarray}
\noindent
These are the desired contributions to the imaginary parts of the polarization tensor
of graphene depending on temperature and on the chemical potential.

\subsection{Real part of conductivity at any temperature}

In accordance with Eq.~(\ref{eq17}), the total imaginary part of the polarization
tensor in the two lowest perturbation orders is obtained by a summation of
Eq.~(\ref{eq19}) and  (\ref{eq26}). In making this summation we take into account
that
\begin{equation}
\frac{1}{2}-\frac{1}{e^y+1}=\frac{1}{2}\,{\rm tanh}\frac{y}{2}
\label{eq27}
\end{equation}
\noindent
and obtain
\begin{eqnarray}
&&
{\rm Im}\Pi_{00}{\ar}=\frac{\pi\alpha\hbar ck^2}{\omega}\,
\theta(\ho\eta-\Delta)
\sum_{\kappa=\pm 1}\left\{
\frac{(\ho)^2+\Delta^2}{2(\ho)^2}
{\rm tanh}\frac{\ho+2\kappa\mu}{4k_BT}\right.
\nonumber \\
&&~~~
+\frac{v_F^2k^2}{4\omega^2}\frac{(\ho)^2+3\Delta^2}{(\ho)^2}   \left[
{\rm tanh}\frac{\ho+2\kappa\mu}{4k_BT}
-\left.
\frac{(\ho)^2-\Delta^2}{4(k_BT)^2}
{\rm csch}^3\frac{\ho+2\kappa\mu}{2k_BT}
{\rm sinh}^4\frac{\ho+2\kappa\mu}{4k_BT}
\right]\right\},
\nonumber \\
&&
{\rm Im}\Pi{\ar}=-\frac{\pi\alpha\hbar\omega k^2}{c}\,
\theta(\ho\eta-\Delta)
\sum_{\kappa=\pm 1}\left\{
\frac{(\ho)^2+\Delta^2}{2(\ho)^2}
{\rm tanh}\frac{\ho+2\kappa\mu}{4k_BT}\right.
\label{eq28} \\
&&~~~
-\frac{v_F^2k^2}{4\omega^2}\frac{(\ho)^2-\Delta^2}{(\ho)^2}   \left[
{\rm tanh}\frac{\ho+2\kappa\mu}{4k_BT}
-\left.
\frac{3(\ho)^2+\Delta^2}{4(k_BT)^2}
{\rm csch}^3\frac{\ho+2\kappa\mu}{2k_BT}
{\rm sinh}^4\frac{\ho+2\kappa\mu}{4k_BT}
\right]\right\}.
\nonumber
\end{eqnarray}

Substituting Eq.~(\ref{eq28}) in Eq.~(\ref{eq16}), one obtains the final result
for the conductivities of graphene with arbitrary mass gap and chemical potential
at any temperature
\begin{eqnarray}
&&
{\rm Re}\sila=\sigma_0
\theta(\ho\eta-\Delta)
\sum_{\kappa=\pm 1}\left\{
\frac{(\ho)^2+\Delta^2}{2(\ho)^2}
{\rm tanh}\frac{\ho+2\kappa\mu}{4k_BT}\right.
\nonumber \\
&&~~~
+\frac{v_F^2k^2}{4\omega^2}\frac{(\ho)^2+3\Delta^2}{(\ho)^2}   \left[
{\rm tanh}\frac{\ho+2\kappa\mu}{4k_BT}
-\left.
\frac{(\ho)^2-\Delta^2}{4(k_BT)^2}
{\rm csch}^3\frac{\ho+2\kappa\mu}{2k_BT}
{\rm sinh}^4\frac{\ho+2\kappa\mu}{4k_BT}
\right]\right\},
\nonumber \\
&&
{\rm Re}\sita=\sigma_0
\theta(\ho\eta-\Delta)
\sum_{\kappa=\pm 1}\left\{
\frac{(\ho)^2+\Delta^2}{2(\ho)^2}
{\rm tanh}\frac{\ho+2\kappa\mu}{4k_BT}\right.
\label{eq29} \\
&&~~~
-\frac{v_F^2k^2}{4\omega^2}\frac{(\ho)^2-\Delta^2}{(\ho)^2}   \left[
{\rm tanh}\frac{\ho+2\kappa\mu}{4k_BT}
-\left.
\frac{3(\ho)^2+\Delta^2}{4(k_BT)^2}
{\rm csch}^3\frac{\ho+2\kappa\mu}{2k_BT}
{\rm sinh}^4\frac{\ho+2\kappa\mu}{4k_BT}
\right]\right\}.
\nonumber
\end{eqnarray}

The conductivities (\ref{eq29}) are nonlocal, but the corrections to local results
are of order $v_F^2/c^2\sim 10^{-5}$.
The main contributions to the conductivity of graphene are obtained in the
local limit
\begin{eqnarray}
&&
{\rm Re}\sil(\omega,0,T,\Delta,\mu)=
{\rm Re}\sit(\omega,0,T,\Delta,\mu)=\sigma_0
\theta(\ho\eta-\Delta)
\nonumber \\
&&~~~~
\times
\frac{(\ho)^2+\Delta^2}{2(\ho)^2}\left(
{\rm tanh}\frac{\ho+2\mu}{4k_BT}+{\rm tanh}\frac{\ho-2\mu}{4k_BT}\right).
\label{eq30}
\end{eqnarray}
\noindent
At high frequencies satisfying the condition $\ho\gg2\mu$ Eq.~(\ref{eq30})
leads to
\begin{eqnarray}
&&
{\rm Re}\sigma_{\|(\bot)}(\omega,0,T,\Delta,\mu)=\sigma_0
\theta(\ho\eta-\Delta)\frac{(\ho)^2+\Delta^2}{(\ho)^2}
{\rm tanh}\frac{\ho}{4k_BT}
\nonumber \\
&&~~~~~~~
\times\left[
1+\left(\frac{\mu}{2k_BT}{\rm sech}\frac{\ho}{4k_BT}\right)^2\right].
\label{eq31}
\end{eqnarray}
\noindent
Alternatively, at low frequencies $\Delta<\ho\ll2\mu$ one has from Eq.~(\ref{eq30})
\begin{equation}
{\rm Re}\sigma_{\|(\bot)}(\omega,0,T,\Delta,\mu)=\sigma_0
\theta(\ho\eta-\Delta)\frac{(\ho)^2+\Delta^2}{(\ho)^2}
\frac{\ho}{4k_BT}{\rm sech}^2\frac{\mu}{2k_BT}.
\label{eq32}
\end{equation}

Note that the same dependence of the conductivity of graphene on the frequency,
temperature and chemical potential, as presented inside the round brackets in
Eq.~(\ref{eq30}), was obtained in Ref.~\cite{11} within the approximate approach
starting from the Kubo formula (see also Ref.~\cite{68} where this dependence was
found for the case $\mu=0$). In Ref.~\cite{69} Eq.~(\ref{eq30}) is contained
for the case of
gapless graphene, $\Delta=0$, and interpreted as originating from
the interband transitions.

The analytic result (\ref{eq30}), although simple, suggests rich set of dependences
of the real part of conductivity of graphene on the temperature and chemical
potential for various relationships between $\mu$ and $\Delta$.
Before making numerical computations, we briefly discuss the typical values
of $\Delta$ and $\mu$. For a free-standing or deposited on a substrate graphene
sheet an upper bound for the mass gap was estimated as $\Delta<0.2\,$eV
\cite{3,39}. The value of the chemical potential at zero temperature can be
expressed via the doping concentration $n$ as \cite{70}
\begin{equation}
\mu=\hbar v_F\sqrt{\pi n}.
\label{eq33}
\end{equation}
\noindent
For example, for the doping concentrations $n\approx 7.5\times 10^{11}$ and
$2\times 10^{13}\,\mbox{cm}^{-2}$ one obtains $\mu=0.1$ and 0.5\,eV,
respectively. In the experiment on measuring the Casimir interaction between
a Au-coated sphere and a graphene-coated SiO${}_2$ layer on the top of a Si
plate it was found \cite{71} that $ n\approx1.2\times 10^{10}\,\mbox{cm}^{-2}$,
 which leads from Eq.~(\ref{eq33}) to the upper bound of
$\mu=0.02\,$eV. This experiment was performed in high vacuum.
In the measurement of the optical conductivity of graphene  on top of
a SiO${}_2$ substrate \cite{69} a representative value of $\mu=0.1\,$eV
has been used.
Note also that we do not consider the disorder effects in the form of, e.g.,
the charged puddles violating translational invariance, which may arise due to
substrate and environment in addition to a  relatively homogeneous doping.

In Fig.~\ref{fg1} we plot the real part of the conductivity of graphene
with the mass gap $\Delta=0.02\,$eV normalized to $\sigma_0$ as a function
of frequency at $T=10\,$K, 100\,K, and 300\,K (lines 1, 2, and 3,
respectively) for (a) any value of the chemical potential satisfying
a condition $\mu< 0.01\,$eV and (b) for $2\mu=0.07\,$eV.
Note that the positions of lines in Fig.~\ref{fg1}(a) do not depend on
the specific value of $\mu$ in the indicated interval.
In both Figs.~\ref{fg1}(a) and \ref{fg1}(b)
the real part of conductivity vanishes for $\ho<\Delta=0.02\,$eV.
As is seen in Figs.~\ref{fg1}(a) and \ref{fg1}(b), the character of
${\rm Re}\sigma_{\|(\bot)}$ as a function of frequency changes qualitatively
depending on whether $2\mu<\Delta$ or  $2\mu>\Delta$. Under these conditions
the real part of conductivity of graphene behaves quite differently with
decreasing temperature (see below for the case $T=0$).
With increasing $\omega$ the real part of conductivity in all cases goes to the
universal conductivity $\sigma_0$.

To illustrate the dependence of the real part of conductivity on the chemical
potential, we introduce the quantity
\begin{equation}
\eta_{\mu}(\omega,T,\mu)=
\frac{{\rm Re}\sigma_{\|(\bot)}(\omega,0,T,\Delta,\mu)}{{\rm Re}\sigma_{\|(\bot)}
(\omega,0,T,\Delta,0)},
\label{eq34}
\end{equation}
\noindent
which does not depend on the mass gap $\Delta$.
In Fig.~\ref{fg2} we plot the quantity $\eta_{\mu}$ as a function of chemical
potential at $T=10\,$K, 100\,K, and 300\,K
(lines 1, 2, and 3, respectively) for the three values of
frequency (a) $\ho=0.01\,$eV,
(b) $\ho=0.05\,$eV, and (c) $\ho=0.1\,$eV.
The respective possible values of the mass gap, for which the real part of
conductivity does not vanish, satisfy the inequalities
(a) $\Delta<0.01\,$eV,
(b) $\Delta<0.05\,$eV, and (c) $\Delta<0.1\,$eV.
As is seen in Fig.~\ref{fg2}, the impact of chemical potential essentially
depends on both the temperature and frequency taking into account also that
the latter restricts the allowed values of the mass gap.

Now we consider the real part of the conductivity of graphene at zero
temperature. For this purpose we take the limit $T\to 0$ in Eq.~(\ref{eq29}).
It is easily seen that the last terms in both ${\rm Re}\sil$ and
${\rm Re}\sit$, which contain
${\rm sinh}^4[(\ho+2\kappa\mu)/(4k_BT)]$, go to zero when $T$ goes to zero
independently of the sign of $\kappa$.
As to the function
${\rm tanh}[(\ho+2\kappa\mu)/(4k_BT)]$,  it goes to unity
when $T$ goes to zero if $\kappa=+1$, and also if $\kappa=-1$ and $\ho>2\mu$.
If $\kappa=-1$ and $\ho<2\mu$ this function goes to minus unity with
vanishing temperature. Taking all this into account, for $\ho>2\mu$
one obtains
\begin{eqnarray}
&&
{\rm Re}\sil(\omega,k,0,\Delta,\mu)=\sigma_0
\theta(\ho\eta-\Delta)
\nonumber \\
&&~~
\times\left[
\frac{(\ho)^2+\Delta^2}{(\ho)^2}
+\frac{v_F^2k^2}{2\omega^2}\frac{(\ho)^2+3\Delta^2}{(\ho)^2}
\right],
\nonumber \\
&&
{\rm Re}\sit(\omega,k,0,\Delta,\mu)=\sigma_0
\theta(\ho\eta-\Delta)
\label{eq35} \\
&&~~
\times\left[
\frac{(\ho)^2+\Delta^2}{(\ho)^2}
-\frac{v_F^2k^2}{2\omega^2}\frac{(\ho)^2-\Delta^2}{(\ho)^2}
\right].
\nonumber
\end{eqnarray}
\noindent
These expressions are determined by only ${\rm Im}\Pi_{00}^{(0)}$
and ${\rm Im}\Pi^{(0)}$ defined in Eq.~(\ref{eq19}).

In the local limit for $\ho>2\mu$ the real part of the conductivity
of graphene at zero temperature takes an especially simple form
\begin{equation}
{\rm Re}\sil(\omega,0,0,\Delta,\mu)=
{\rm Re}\sit(\omega,0,0,\Delta,\mu)=\sigma_0
\theta(\ho\eta-\Delta)
\frac{(\ho)^2+\Delta^2}{(\ho)^2}.
\label{eq36}
\end{equation}

Alternatively, for $\ho<2\mu$ in the limiting case $T\to 0$
Eq.~(\ref{eq29}) leads to the vanishing real part of the conductivities
of graphene
\begin{equation}
{\rm Re}\sil(\omega,0,0,\Delta,\mu)=
{\rm Re}\sit(\omega,0,0,\Delta,\mu)=0.
\label{eq37}
\end{equation}
\noindent
This result is determined by the canceled contributions from
${\rm Im}\Pi_{00}^{(0)}$, ${\rm Im}\Pi^{(0)}$, on the one hand, and
${\rm Im}\Pi_{00}^{(1)}$, ${\rm Im}\Pi^{(1)}$, on the other hand.
For a gapped graphene with zero chemical potential Eq.~(\ref{eq36})
is presented in Ref.~\cite{12} (vanishing of the conductivity of
gapped graphene at low frequencies was also noticed in Refs.~\cite{22,24}).
In Ref.~\cite{61} Eqs.~(\ref{eq35}) and (\ref{eq36}) were derived for the
case of gapped but undoped ($\mu=0$) graphene using the formalism of the
polarization tensor.

According to the above results (\ref{eq35})--(\ref{eq37}), which are valid
for both gapped and doped graphene, the real parts of graphene conductivities
vanish in the frequency interval
$[0,\max\{\Delta,2\mu\}/\hbar]$.
In doing so, the magnitude of ${\rm Re}\sigma_{\|(\bot)}$ is always determined
by the magnitude of the mass gap $\Delta$. In this sense there is no full symmetry
between the cases of gapped, but undoped, and gapless, but doped, graphene
mentioned in Sec.~I if to identify $\Delta$ with $2E_F$ (recall that at $T=0$
the chemical potential $\mu$ is just the Fermi energy of the nonrelativistic
electrons \cite{70}). Note that vanishing of the conductivity of graphene in
the interval $[0,\max\{\Delta,2\mu\}/\hbar]$ was discussed  in Ref.~\cite{22}.

The tendency toward vanishing of the real part of graphene conductivity in
the interval mentioned above can be traced in Fig.~\ref{fg1}(b), where with
decreasing temperature (line 1 is plotted for $T=10\,$K) the right boundary
of the range, where  ${\rm Re}\sigma_{\|(\bot)}$ is zero approaches to
$2\mu=0.07\,$eV. In Fig.~\ref{fg1}(a), where $2\mu<\Delta$, this range is
determined by $\Delta$ at all temperatures.

In Fig.~\ref{fg3} we illustrate the dependence ${\rm Re}\sigma_{\|(\bot)}$,
normalized to the universal conductivity $\sigma_0$, on the dimensionless
parameter $\ho/\Delta$. Figure~\ref{fg3}(a) is plotted for the case
$2\mu<\Delta$. Here, the frequency region where ${\rm Re}\sigma_{\|(\bot)}=0$
is $[0,\Delta/\hbar)$, i.e., is determined by the mass-gap parameter.
In Fig.~\ref{fg3}(b) the case $\mu=\Delta$ is considered. Now the
region where ${\rm Re}\sigma_{\|(\bot)}=0$
is $[0,2\mu/\hbar)$, i.e., is determined by the value of chemical potential.
In both Figs.~\ref{fg3}(a) and \ref{fg3}(b), however, the magnitudes of
${\rm Re}\sigma_{\|(\bot)}$ are determined solely by the value of $\Delta$.
In the limiting case of high frequencies ($\omega\to\infty$) the real part
of the conductivity of graphene goes to the universal conductivity $\sigma_0$,
as it should be on the basis of all previous knowledge.

\section{Impact of chemical potential on imaginary part of conductivity}

According to Eq.~(\ref{eq2}), the imaginary part of conductivity of graphene is
expressed via the real part of polarization tensor
\begin{eqnarray}
&&
{\rm Im}\sila=-\frac{\omega}{4\pi\hbar k^2}{\rm Re}\Pi_{00}{\ar},
\nonumber \\
&&
{\rm Im}\sita=\frac{c^2}{4\pi\hbar\omega k^2}{\rm Re}\Pi{\ar}.
\label{eq39}
\end{eqnarray}
\noindent
In accordance to Eq.~({\ref{eq3})
\begin{eqnarray}
&&
{\rm Re}\Pi_{00}(\omega,k,T,\Delta,\mu)=
{\rm Re}\Pi_{00}^{(0)}(\omega,k,\Delta)+
{\rm Re}\Pi_{00}^{(1)}(\omega,k,T,\Delta,\mu),
\nonumber \\
&&
{\rm Re}\Pi(\omega,k,T,\Delta,\mu)={\rm Re}\Pi^{(0)}(\omega,k,\Delta)+
{\rm Re}\Pi^{(1)}(\omega,k,T,\Delta,\mu).
\label{eq40}
\end{eqnarray}
\noindent
The first contributions on the right-hand side of Eq.~(\ref{eq40}) are
contained in Eqs.~(\ref{eq4}) and  (\ref{eq6}).
They can be identically rewritten in a uniform way for both $\ho\eta<\Delta$
and $\ho\eta\geq\Delta$
\begin{eqnarray}
&&
{\rm Re}\Pi_{00}^{(0)}(\omega,k,\Delta)=
-\frac{2\alpha k^2c}{\omega^2\eta^2}\left\{\Delta-\frac{1}{2}
\ho\eta\left[1+\left(\frac{\Delta}{\ho\eta}\right)^2\right]
\ln\left|\frac{\ho\eta+\Delta}{\ho\eta-\Delta}\right|\right\},
\nonumber \\[2mm]
&&
{\rm Re}\Pi^{(0)}(\omega,k,\Delta)=
\frac{2\alpha k^2}{c}\left\{\Delta-\frac{1}{2}
\ho\eta\left[1+\left(\frac{\Delta}{\ho\eta}\right)^2\right]
\ln\left|\frac{\ho\eta+\Delta}{\ho\eta-\Delta}\right|
\right\}.
\label{eq41}
\end{eqnarray}
\noindent
Thus, the real parts of $\Pi_{00}^{(0)}$ and $\Pi^{(0)}$  are not equal to
zero at all frequencies.

Now we take into account that due to Eq.~(\ref{eq20}) the nonlocal corrections
to the conductivity of graphene are very small as compared to the main, local,
contribution. Because of this, here we make all calculations in the lowest order
of the small parameter $v_F^2k^2/\omega^2$.
Thus, expanding Eq.~(\ref{eq41}) to the lowest order in this parameter,
one obtains
\begin{eqnarray}
&&
{\rm Re}\Pi_{00}^{(0)}(\omega,k,\Delta)=
-\frac{2\alpha\hbar c k^2}{\omega}\left[\frac{\Delta}{\ho}-
\frac{(\ho)^2+\Delta^2}{2(\ho)^2}
\ln\left|\frac{\ho+\Delta}{\ho-\Delta}\right|
\right],
\nonumber \\[2mm]
&&
{\rm Re}\Pi^{(0)}(\omega,k,\Delta)=
\frac{2\alpha \ho k^2}{c}\left[\frac{\Delta}{\ho}-
\frac{(\ho)^2+\Delta^2}{2(\ho)^2}
\ln\left|\frac{\ho+\Delta}{\ho-\Delta}\right|
\right].
\label{eq42}
\end{eqnarray}

The second contributions on the right-hand side of Eq.~(\ref{eq40}) are
given by Eqs.~(\ref{eq10})--(\ref{eq14}). They are also not equal to
zero at all frequencies. We start from obtaining the main terms in the
expansion of ${\rm Re}\Pi_{00}^{(1)}$ and ${\rm Re}\Pi^{(1)}$ in the powers
of $v_F^2k^2/\omega^2$.

\subsection{Real parts of contributions to the polarization tensor
depending on chemical potential}

We consider first the frequency region $\ho\eta<\Delta$ which reduces to
$\ho<\Delta$ in the lowest order of our small parameter. In this region
${\rm Re}\Pi_{00}^{(1)}$ and ${\rm Re}\Pi^{(1)}$ are given by Eq.~(\ref{eq10}).
In the lowest order of the parameter $v_F^2k^2/\omega^2$ one has
\begin{eqnarray}
&&
1-\frac{1}{2\omega\eta}\left[B_1(2cu+\omega)-B_1(2cu-\omega)\right]
\nonumber \\
&&~~
=-\frac{v_F^2k^2}{2\omega^2}\left[1+\frac{(\ho)^2+\Delta^2}{2(\ho)^2}
\left(\frac{\omega}{2cu-\omega}- \frac{\omega}{2cu+\omega}\right)\right],
\nonumber \\
&&
1-\frac{\eta}{2\omega}\left[B_2(2cu+\omega)-B_2(2cu-\omega)\right]
\label{eq43} \\
&&~~
=\frac{v_F^2k^2}{2\omega^2}\left[1+\frac{(\ho)^2+\Delta^2}{2(\ho)^2}
\left(\frac{\omega}{2cu-\omega}- \frac{\omega}{2cu+\omega}\right)\right],
\nonumber
\end{eqnarray}
\noindent
Substituting Eq.~(\ref{eq43}) in  Eq.~(\ref{eq10}) and introducing the
new integration variable $t=2vu/\omega$, we arrive at
\begin{eqnarray}
&&
{\rm Re}\Pi_{00}^{(1)}{\ar}=-\frac{2\alpha\hbar c k^2}{\omega}
\int_{\Delta/(\ho)}^{\infty}\!\!dt\sum_{\kappa=\pm 1}
\frac{1}{e^{\frac{\ho t+2\kappa\mu}{2k_BT}}+1}
\nonumber \\
&&~~~~
\times\left[1+\frac{(\ho)^2+\Delta^2}{(\ho)^2}\,\frac{1}{t^2-1}\right],
\nonumber \\
&&
{\rm Re}\Pi^{(1)}{\ar}=\frac{2\alpha\hbar\omega k^2}{c}
\int_{\Delta/(\ho)}^{\infty}\!\!dt\sum_{\kappa=\pm 1}
\frac{1}{e^{\frac{\ho t+2\kappa\mu}{2k_BT}}+1}
\nonumber \\
&&~~~~
\times\left[1+\frac{(\ho)^2+\Delta^2}{(\ho)^2}\,\frac{1}{t^2-1}\right].
\label{eq44}
\end{eqnarray}
\noindent
Note that in the frequency region under consideration $\Delta/(\ho)>1$.

We are coming now to the case $\ho\eta\geq\Delta$ which reduces to
$\ho\geq\Delta$ in the lowest perturbation order.
Taking into account that in the lowest order the quantities
${\rm Re}\Pi_{00}^{(1)}$ and ${\rm Re}\Pi^{(1)}$
differ by only a factor in front of the integral [see Eq.~(\ref{eq44})]
we can restrict ourselves by considering  ${\rm Re}\Pi_{00}^{(1)}$
defined in Eqs.~(\ref{eq11}) and (\ref{eq12}).
Then, the sum of the integrals defined in Eq.~(\ref{eq12}) is given by
\begin{eqnarray}
&&
I_1+I_2+I_3=-\frac{v_F^2 k^2}{2\omega^2}
\int_{\Delta/(2\hbar c)}^{\infty}\!\!du\sum_{\kappa=\pm 1}
\frac{1}{e^{\beta u+\frac{\kappa\mu}{k_BT}}+1}
\nonumber \\
&&~~~~
\times\left[1+\frac{(\ho)^2+\Delta^2}{2(\ho)^2}\,
\left(\frac{\omega}{2cu-\omega}- \frac{\omega}{2cu+\omega}\right)
\right]
\label{eq45} \\
&&~
-\frac{1}{2}
\int_{u^{(-)}}^{u^{(+)}}\!\!du\sum_{\kappa=\pm 1}
\frac{1}{e^{\beta u+\frac{\kappa\mu}{k_BT}}+1}
\left\{\frac{2cu-\omega}{\omega}+\frac{v_F^2k^2}{2\omega^2}\left[
\frac{2cu-\omega}{\omega}-\frac{(2-A_0)\omega}{2cu-\omega}
\right]\right\}.
\nonumber
\end{eqnarray}

It can be easily seen that the second integral on the right-hand side of
Eq.~(\ref{eq45}) is of the order of $v_F^3k^3/\omega^3$, i.e., should be
omitted in our calculation. Then, substituting Eq.~(\ref{eq45})  in
Eq.~(\ref{eq11}) and introducing the integration variable $t=2cu/\omega$,
one obtains the real part of the $\mu$-dependent contribution to the
polarization tensor
\begin{eqnarray}
&&
{\rm Re}\Pi_{00}^{(1)}{\ar}=-\frac{2\alpha\hbar c k^2}{\omega}
\int_{\Delta/(\ho)}^{\infty}\!\!dt\sum_{\kappa=\pm 1}
\frac{1}{e^{\frac{\ho t+2\kappa\mu}{2k_BT}}+1}
\nonumber \\
&&~~~~
\times\left[1+\frac{(\ho)^2+\Delta^2}{(\ho)^2}\,\frac{1}{t^2-1}\right].
\label{eq46}
\end{eqnarray}
\noindent
The same expression, but with the factor $2\alpha\hbar\omega k^2/c$ in
front of the summation sign, is obtained for ${\rm Re}\Pi^{(1)}$.
It is seen that Eq.~(\ref{eq46}) is identical in form to
Eq.~(\ref{eq44}), but the singular point $t=1$ in Eq.~(\ref{eq46})
is inside the integration interval, i.e., the integral in this case is the
improper one. The integrals in Eqs.~ (\ref{eq44}) and (\ref{eq46})
can be calculated either numerically or analytically in some asymptotic
regions (see below).

\subsection{Imaginary part of conductivity at any temperature}

The total real part of the polarization tensor in the lowest perturbation
order is obtained by the summation of Eqs.~ (\ref{eq42}) and (\ref{eq44})
\begin{eqnarray}
&&
{\rm Re}\Pi_{00}{\ar}=-\frac{2\alpha\hbar c k^2}{\omega}
\left\{\vphantom{\frac{1}{e^{\frac{\ho t+2\kappa\mu}{2k_BT}}+1}}
\frac{\Delta}{\ho}-
\frac{(\ho)^2+\Delta^2}{2(\ho)^2}
\ln\left|\frac{\ho+\Delta}{\ho-\Delta}\right|
\right.
\nonumber \\
&&~~
+\left.
\int_{\Delta/(\ho)}^{\infty}\!\!dt\sum_{\kappa=\pm 1}
\frac{1}{e^{\frac{\ho t+2\kappa\mu}{2k_BT}}+1}
\left[1+\frac{(\ho)^2+\Delta^2}{(\ho)^2}\,\frac{1}{t^2-1}\right]
\right\}.
\label{eq47}
\end{eqnarray}
\noindent
The same expression, but with the factor $2\alpha\hbar\omega k^2/c$ in
front of the figure brackets, is obtained for ${\rm Re}\Pi\ar$.

Substituting these expressions in Eq.~ (\ref{eq39}), for the imaginary parts
of the conductivities of graphene in the local limit one finds
\begin{eqnarray}
&&
{\rm Im}\sil(\omega,0,T,\Delta,\mu)={\rm Im}\sit(\omega,0,T,\Delta,\mu)
\label{eq48} \\
&&~
=\frac{\sigma_0}{\pi}\left[
\frac{2\Delta}{\ho}-
\frac{(\ho)^2+\Delta^2}{(\ho)^2}
\ln\left|\frac{\ho+\Delta}{\ho-\Delta}\right|
+Y(\omega,T,\Delta,\mu)\right]
\nonumber
\end{eqnarray}
\noindent
where
\begin{eqnarray}
&&
Y(\omega,T,\Delta,\mu)=2
\int_{\Delta/(\ho)}^{\infty}\!\!dt\sum_{\kappa=\pm 1}
\frac{1}{e^{\frac{\ho t+2\kappa\mu}{2k_BT}}+1}
\nonumber \\
&&~~~~~~
\times
\left[1+\frac{(\ho)^2+\Delta^2}{(\ho)^2}\,\frac{1}{t^2-1}\right].
\label{eq49}
\end{eqnarray}
\noindent
This result is valid at all frequencies, i.e., for both $\ho<\Delta$ and
$\ho\geq\Delta$.

In Fig.~\ref{fg4} we present the results of numerical computations of
${\rm Im}\sigma_{\|(\bot)}$ by Eq.~(\ref{eq48}) as a function of chemical
potential for $T=300\,$K, $\Delta=0.1\,$eV and three values of frequency
$\ho=0.01$, 0.05, and 0.099\,eV (lines 1, 2, and 3, respectively).
As is seen in Fig.~\ref{fg4}, the imaginary part of the conductivity increases
monotonously with increasing $\mu$. This increase is more rapid at lower
frequencies. In so doing, the imaginary part of graphene conductivity may take
both negative and positive values.

Now we consider the asymptotic limit of the imaginary part of conductivity
(\ref{eq48}) at low temperature $k_BT\ll\mu$. In this case the integral
in Eq.~(\ref{eq49}) takes different values for $\Delta<2\mu$ and $\Delta>2\mu$.
Thus, under the condition  $\Delta<2\mu$  and for the frequencies $\ho\ll 2\mu$
we arrive at (see the Appendix)
\begin{eqnarray}
&&
Y(\omega,T,\Delta,\mu)=\frac{4\mu}{\ho}-\frac{2\Delta}{\ho}-4\ln2\,\frac{k_BT}{\ho}
\nonumber \\
&&~~~
-\frac{(\ho)^2+\Delta^2}{(\ho)^2}\left(\ln\frac{2\mu+\ho}{2\mu-\ho}-
\ln\left|\frac{\ho+\Delta}{\ho-\Delta}\right|\right).
\label{eq50}
\end{eqnarray}

Substituting Eq.~(\ref{eq50}) in Eq.~(\ref{eq48}), one obtains
\begin{equation}
{\rm Im}\sigma_{\|(\bot)}(\omega,0,T,\Delta,\mu)=\frac{\sigma_0}{\pi}\left[
\frac{4\mu}{\ho}-4\ln2\,\frac{k_BT}{\ho}
-\frac{(\ho)^2+\Delta^2}{(\ho)^2}\ln\frac{2\mu+\ho}{2\mu-\ho}\right].
\label{eq51}
\end{equation}
\noindent
The first Drude-like term on the right-hand side of this equation
corresponds to the imaginary part of graphene conductivity at low temperature obtained
in Ref.~\cite{19} for the case $\Delta=0$ and interpreted as originating from the
intraband transitions.

Under the same condition  $\Delta<2\mu$, but for the frequencies $\ho\gg 2\mu$
the derivation presented in the Appendix leads to
\begin{eqnarray}
&&
Y(\omega,T,\Delta,\mu)=\frac{4\mu}{\ho}-\frac{2\Delta}{\ho}+4\ln2\,\frac{k_BT}{\ho}
\frac{\Delta^2}{(\ho)^2}
\nonumber \\
&&~~~
-\frac{(\ho)^2+\Delta^2}{(\ho)^2}\left(\ln\frac{\ho+2\mu}{\ho-2\mu}-
\ln\frac{\ho+\Delta}{\ho-\Delta}\right).
\label{eq52}
\end{eqnarray}
\noindent
After substitution of this equation  in Eq.~(\ref{eq48}), we find
\begin{equation}
{\rm Im}\sigma_{\|(\bot)}(\omega,0,T,\Delta,\mu)=\frac{\sigma_0}{\pi}\left[
\frac{4\mu}{\ho}+4\ln2\,\frac{k_BT}{\ho}\frac{\Delta^2}{(\ho)^2}
-\frac{(\ho)^2+\Delta^2}{(\ho)^2}\ln\frac{\ho+2\mu}{\ho-2\mu}\right].
\label{eq53}
\end{equation}

If the opposite condition $\Delta>2\mu$ is satisfied, calculations made in the
Appendix lead to the following result valid at all frequencies:
\begin{equation}
Y(\omega,T,\Delta,\mu)=8\ln2\frac{\Delta^2}{\Delta^2-(\ho)^2}
\frac{k_BT}{\ho}e^{-\frac{\Delta}{k_BT}}{\rm cosh}\frac{\mu}{k_BT}.
\label{eq54}
\end{equation}
\noindent
{}From Eqs.~(\ref{eq54}) and (\ref{eq48}) for the imaginary part of
the conductivity it follows
\begin{eqnarray}
&&
{\rm Im}\sigma_{\|(\bot)}(\omega,0,T,\Delta,\mu)=\frac{\sigma_0}{\pi}\left[
\frac{2\Delta}{\ho} +8\ln2\frac{\Delta^2}{\Delta^2-(\ho)^2}
\frac{k_BT}{\ho}e^{-\frac{\Delta}{k_BT}}{\rm cosh}\frac{\mu}{k_BT}
\right.
\nonumber \\
&&~~~~~~~~\left.
-\frac{(\ho)^2+\Delta^2}{(\ho)^2}
\ln\left|\frac{\ho+\Delta}{\ho-\Delta}\right|\right].
\label{eq55}
\end{eqnarray}

As is seen in Eqs.~(\ref{eq51}) and (\ref{eq53}), in the case $\Delta<2\mu$
there is the linear in temperature term at low temperatures. This term is
much smaller in the frequency region $\ho\gg 2\mu$ as compared to the frequency
region  $\ho\ll 2\mu$. In the case $\Delta>2\mu$ the thermal effect in the
imaginary part of the conductivity of graphene is exponentially small
[see Eq.~(\ref{eq55})].

Now we consider the case of zero temperature $T=0$. In the case $\Delta>2\mu$
both exponents in Eq.~(\ref{eq49}) have the positive powers and go to infinity
when the temperature vanishes. Thus, $Y=0$ and, according to Eq.~(\ref{eq48}),
the imaginary part of the conductivity of graphene is given by
\begin{equation}
{\rm Im}\sigma_{\|(\bot)}(\omega,0,0,\Delta,\mu)=\frac{\sigma_0}{\pi}\left[
\frac{2\Delta}{\ho}
-\frac{(\ho)^2+\Delta^2}{(\ho)^2}
\ln\left|\frac{\ho+\Delta}{\ho-\Delta}\right|\right].
\label{eq56}
\end{equation}
\noindent
In this case it is determined by only ${\rm Re}\Pi_{00}^{(0)}$ and
${\rm Re}\Pi^{(0)}$. There is no explicit dependence of ${\rm Im}\sigma_{\|(\bot)}$
on the chemical potential when $\Delta>2\mu$. Note that Eq.~(\ref{eq56}) is
presented in Ref.~\cite{12} for a gapped but undoped graphene.

If $\Delta<2\mu$, the power of the exponent with $\kappa=-1$ is negative for
$t<2\mu/(\ho)$. Then, in the limit of zero temperature, the exponent-containing
fraction goes to unity and Eq.~(\ref{eq49}) results in
\begin{equation}
Y(\omega,0,\Delta,\mu)=2
\int_{\Delta/(\ho)}^{2\mu/(\ho)}\!\!dt
\left[1+\frac{(\ho)^2+\Delta^2}{(\ho)^2}\,\frac{1}{t^2-1}\right].
\label{eq57}
\end{equation}

After the integration we obtain
\begin{eqnarray}
&&
Y(\omega,0,\Delta,\mu)=\frac{4\mu}{\ho}-\frac{2\Delta}{\ho}
\nonumber \\
&&~~~
-\frac{(\ho)^2+\Delta^2}{(\ho)^2}\left(
\ln\left|\frac{\ho+2\mu}{\ho-2\mu}\right|-
\ln\left|\frac{\ho+\Delta}{\ho-\Delta}\right|\right).
\label{eq58}
\end{eqnarray}
\noindent
Substituting Eq.~(\ref{eq58}) in Eq.~(\ref{eq48}), we arrive at
\begin{equation}
{\rm Im}\sigma_{\|(\bot)}(\omega,0,0,\Delta,\mu)=\frac{\sigma_0}{\pi}\left[
\frac{4\mu}{\ho}
-\frac{(\ho)^2+\Delta^2}{(\ho)^2}
\ln\left|\frac{\ho+2\mu}{\ho-2\mu}\right|\right].
\label{eq59}
\end{equation}
\noindent
Here, the imaginary part of graphene conductivity is determined by both
${\rm Re}\Pi_{00}^{(0)}$ and ${\rm Re}\Pi^{(0)}$, on the one hand, and
${\rm Re}\Pi_{00}^{(1)}$ and ${\rm Re}\Pi^{(1)}$, on the other hand.
It is seen that Eq.~(\ref{eq56}) coincides with the limiting case of Eq.~(\ref{eq55})
when $T$ goes to zero. In a similar way, Eq.~(\ref{eq59}) is the limit of
Eq.~(\ref{eq53}) when $T\to 0$, as it should be.
Note that Eqs.~(\ref{eq56}) and (\ref{eq59}) are valid at $T=0$ for any frequency,
whereas the limiting case of Eq.~(\ref{eq53}) is valid  within  the same
frequency region as Eq.~(\ref{eq53}), i.e., under the condition $\ho\gg2\mu$.

Slightly simpler asymptotic formulas can be obtained also from
Eqs.~(\ref{eq56}) and (\ref{eq59}) under some additional conditions.
 Thus, if $\Delta>2\mu$  and $\ho\ll\Delta$ Eq.~(\ref{eq56}) leads to
\begin{equation}
{\rm Im}\sigma_{\|(\bot)}(\omega,0,0,\Delta,\mu)=-\frac{8\sigma_0}{3\pi}
\frac{\ho}{\Delta}.
\label{eq60}
\end{equation}
\noindent
If $\Delta<2\mu$  and $\ho\ll 2\mu$, Eq.~(\ref{eq59}) results in
\begin{equation}
{\rm Im}\sigma_{\|(\bot)}(\omega,0,0,\Delta,\mu)=\frac{4\mu\sigma_0}{\pi\ho}
\left[1-\left(\frac{\Delta}{2\mu}\right)^2\right].
\label{eq61}
\end{equation}

Now we present the results of numerical computations using
Eqs.~(\ref{eq56}) and (\ref{eq59}). In Fig.~\ref{fg5} we plot the imaginary
part of the conductivity of graphene normalized to the universal conductivity
$\sigma_0$ as a function of the dimensionless parameter $\ho/\Delta$ for (a)
$\Delta>2\mu$ and (b) $\Delta=\mu$. In Fig.~\ref{fg5}(a) computations have been
performed by using Eq.~(\ref{eq56}). In this case the value of
${\rm Im}\sigma_{\|(\bot)}$ does not depend on a specific value of the
chemical potential $\mu$. It is important only that $2\mu<\Delta$.
At $\ho=\Delta$ the imaginary part of the graphene conductivity goes to minus
infinity. For $\ho/\Delta\ll 1$ the imaginary part of conductivity goes to
zero by the linear law [see Eq.~(\ref{eq60})].
In Fig.~\ref{fg5}(b) computations have been
performed by using Eq.~(\ref{eq59}).
Here, the value of the chemical potential $\mu=\Delta$ was chosen, i.e.,
$\Delta<2\mu$. In this case the value of
${\rm Im}\sigma_{\|(\bot)}$ depends essentially on the specific value of $\mu$.
For $\ho/\Delta\ll 1$ the imaginary part of graphene conductivity goes to
infinity in accordance with Eq.~(\ref{eq61}). At $\ho=2\mu=2\Delta$, i.e.,
at $\ho/\Delta=2$, ${\rm Im}\sigma_{\|(\bot)}$ goes to minus
infinity.

\section{Conclusions and discussion}

In this paper, we have developed the complete theory of electrical
conductivity of graphene with taken into account mass-gap parameter and
chemical potential at arbitrary temperature. This was done in the framework
of the Dirac model starting from the first principles of quantum electrodynamics
at nonzero temperature. The main quantity, used to investigate the
conductivity properties of graphene, is the polarization tensor in
(2+1)-dimensional space-time. For a gapped graphene with zero chemical
potential it was initially found only at the pure imaginary
Matsubara frequencies
in Refs.~\cite{39,40} and in the other form, allowing an analytic
continuation to the entire plane of complex frequencies, in Ref.~\cite{52}.
The polarization tensor of gapped graphene with nonzero chemical
potential was presented in Ref.~\cite{62} along the imaginary frequency
axis. Here, we present the analytic continuation of this tensor to the
real frequency axis, which is required for rigorous investigation of the
conductivity of graphene. Specifically, we demonstrate that for nonzero
chemical potential the temperature-dependent contribution to the
polarization tensor at real frequencies does not vanish in the limit of
zero temperature and, generally speaking, results in some additional
terms depending on both the mass gap and chemical potential.

In the framework of this formalism, the real parts of the in-plane and
out-of-plane conductivities of graphene have been found both in the
local approximation and with taken into account corrections due to
nonlocality. The latter were shown to be of the order of $10^{-5}$ in
relation to the main contribution. It was demonstrated that the real
parts of the graphene conductivities are not equal to zero only for
frequencies exceeding the mass-gap parameter. In this frequency region
we have obtained very simple analytic results for the real parts of
the conductivity of graphene at both zero and nonzero temperature.
In the limit of large frequencies  the real part of graphene
 conductivity goes to the universal conductivity.
 In several specific cases our results reproduce
 and generalize the ones obtained earlier
 in the literature using various approximate and phenomenological approaches.
We have also performed numerical computations illustrating the dependence
of the conductivity of graphene on the chemical potential.

Furthermore, the imaginary part of the conductivity of graphene has been
investigated. It was shown to be nontrivial at all frequencies.
In the local limit, the exact expression for the imaginary part of
conductivity was obtained, as well as asymptotic limits of this expression
for different relationships between the mass gap and chemical potential.
In the limiting case of zero temperature, simple analytic results for the
imaginary parts of the graphene conductivity were derive, which turned out to
be quite different depending on whether the mass gap is smaller or larger
than twice the chemical potential.
Specifically, it was shown that in the latter case the imaginary part of
conductivity goes to minus infinity at the frequency equal to the mass gap.
In the former case, the imaginary part of
conductivity turns to infinity  at zero frequency and
to minus infinity at the frequency equal to twice the chemical potential.
In the limit of large frequencies  the imaginary part of graphene
conductivity goes to zero.
The numerical computations for the imaginary part of
conductivity  have been performed at both nonzero and zero temperature
to illustrate the obtained results.

The obtained rigorous results for the electric conductivity of graphene
can be used as a guide when comparing between different
approximate and phenomenological approaches and in the confrontation
between experiment and theory.

\section*{Acknowledgments}
The work of V.M.M. was partially supported by the Russian
Government
Program of Competitive Growth of Kazan Federal University.
\appendix
\section{}
\setcounter{equation}{0}
\renewcommand{\theequation}{A\arabic{equation}}
Here, we derive the asymptotic expressions (\ref{eq50}), (\ref{eq52}) and (\ref{eq54})
for the integral $Y(\omega,T,\Delta,\mu)$ defined in Eq.~(\ref{eq49}).
We consider the limiting case of low temperatures satisfying the condition
$k_BT\ll\mu$. It is easily seen that under this condition the exponent-containing
fractions in Eq.~(\ref{eq48}) with $\kappa=1$ and $\kappa=-1$, but with positive power of
the exponent, can be neglected as compared to the fraction with $\kappa=-1$ and
negative power of exponent. The latter occurs under the conditions $\Delta<2\mu$
and $\hbar\omega t<2\mu$. Taking these conditions into account, the upper integration
limit in Eq.~(\ref{eq48}) can be replaced with $2\mu/(\hbar\omega)$.

Now we expand the remaining exponent-containing
fraction of Eq.~(\ref{eq49}) in the powers of exponent with negative argument
and obtain
\begin{eqnarray}
&&
Y(\omega,T,\Delta,\mu)=2\int_{\Delta/(\hbar\omega)}^{2\mu/(\hbar\omega)}\!\!\!
dt\left[1+\sum_{n=1}^{\infty}(-1)^ne^{\frac{n\hbar\omega t}{2k_BT}-
\frac{n\mu}{k_BT}}\right]
\nonumber \\
&&~~~~~~~
\times\left[1+{\frac{(\hbar\omega)^2+\Delta^2}{(\hbar\omega)^2}}
\frac{1}{t^2-1}\right].
\label{A1}
\end{eqnarray}
\noindent
After performing the
integration of several terms precisely, Eq.~(\ref{A1}) can be rewritten
in the form
\begin{eqnarray}
&&
Y(\omega,T,\Delta,\mu)=
\frac{4\mu}{\hbar\omega}-\frac{2\Delta}{\hbar\omega}-
{\frac{(\hbar\omega)^2+\Delta^2}{(\hbar\omega)^2}}
\left(\ln\left|\frac{2\mu+\hbar\omega}{2\mu-\hbar\omega}\right|
-\ln\left|\frac{\Delta+\hbar\omega}{\Delta-\hbar\omega}\right|\right)
\nonumber \\
&&~~
+2\sum_{n=1}^{\infty} e^{-\frac{n\mu}{k_BT}}(-1)^n
\int_{\Delta/(\hbar\omega)}^{2\mu/(\hbar\omega)}\!\!\!dt
e^{\frac{n\hbar\omega t}{2k_BT}}
\left[1+{\frac{(\hbar\omega)^2+\Delta^2}{(\hbar\omega)^2}}
\frac{1}{t^2-1}\right].
\label{A2}
\end{eqnarray}
\noindent
The larger contributions to the integral in this equation are given by
the larger values of $t$. If we assume in addition that $\hbar\omega\ll2\mu$,
the main contribution to the integral is given by $t\gg 1$. In this case
the quantity in the square brackets can be replaced with unity and after
the integration one obtains
\begin{eqnarray}
&&
Y(\omega,T,\Delta,\mu)=
\frac{4\mu}{\hbar\omega}-\frac{2\Delta}{\hbar\omega}-
{\frac{(\hbar\omega)^2+\Delta^2}{(\hbar\omega)^2}}
\left(\ln\left|\frac{2\mu+\hbar\omega}{2\mu-\hbar\omega}\right|
-\ln\left|\frac{\Delta+\hbar\omega}{\Delta-\hbar\omega}\right|\right)
\nonumber \\
&&~~
+\frac{4k_BT}{\hbar\omega}
\sum_{n=1}^{\infty} \frac{(-1)^n}{n}\left[1-
e^{-\frac{n(2\mu-\Delta)}{2k_BT}}\right].
\label{A3}
\end{eqnarray}
\noindent
Taking into account the above conditions, we find that an exponential term
in the square brackets is much less than unity. Then the summation of the
series results in $-\ln 2$, and from (\ref{A3}) one arrives to
Eq.~(\ref{eq50}).

Now we preserve the condition $\Delta<2\mu$, but, as opposite to the above,
assume that $\hbar\omega\gg 2\mu$.  In this case $t\ll 1$ over the entire
integration region and, thus,
\begin{equation}
\left[1+ {\frac{(\hbar\omega)^2+\Delta^2}{(\hbar\omega)^2}}
\frac{1}{t^2-1}\right]\approx 1-
{\frac{(\hbar\omega)^2+\Delta^2}{(\hbar\omega)^2}}=
-\frac{\Delta^2}{(\hbar\omega)^2}.
\label{A4}
\end{equation}
\noindent
Substituting Eq.~(\ref{A4}) in Eq.~(\ref{A2}) and performing the integration
and summation in the same way as above, one obtains Eq.~(\ref{eq52}).

We continue to consider the case of low temperature, $k_BT\ll\mu$, but under
the opposite condition $\Delta>2\mu$.
In this case the powers of exponents with both $\kappa=1$ and $\kappa=-1$
in Eq.~(\ref{eq49}) are positive. As a result, the main contribution to the
integral is given by the smallest $t\sim\Delta/(\hbar\omega)$.
Consequently, instead of Eq.~(\ref{A4}) one finds
\begin{equation}
\left[1+ {\frac{(\hbar\omega)^2+\Delta^2}{(\hbar\omega)^2}}
\frac{1}{t^2-1}\right]\approx
{\frac{2\Delta^2}{\Delta^2-(\hbar\omega)^2}}.
\label{A5}
\end{equation}
\noindent
Substituting Eq.~(\ref{A5}) in Eq.~(\ref{eq49}) and expanding both fractions in
the powers of exponents with negative arguments, we arrive at
\begin{equation}
Y(\omega,T,\Delta,\mu)=
\frac{4\Delta^2}{\Delta^2-(\hbar\omega)^2}
\int_{\Delta/(\hbar\omega)}^{\infty}\!\!\!
dt\sum_{n=1}^{\infty}(-1)^{n+1}\left(e^{-n\frac{\hbar\omega t+2\mu}{2k_BT}}+
e^{-n\frac{\hbar\omega t-2\mu}{2k_BT}}\right).
\label{A6}
\end{equation}
\noindent
Integration in Eq.~(\ref{A6}) with respect to $t$ leads to
\begin{equation}
Y(\omega,T,\Delta,\mu)=\frac{8k_BT}{\hbar\omega}
\frac{\Delta^2}{\Delta^2-(\hbar\omega)^2}
\sum_{n=1}^{\infty}\frac{(-1)^{n+1}}{n}\left(e^{-n\frac{\Delta+2\mu}{2k_BT}}+
e^{-n\frac{\Delta-2\mu}{2k_BT}}\right).
\label{A7}
\end{equation}
\noindent
Under the above conditions we can restrict ourselves by only the first term
of the series in Eq.~(\ref{A7})  which immediately results in Eq.~(\ref{eq54}).


\newpage
\begin{figure}[b]
\vspace*{-0cm}
\centerline{\hspace*{2.5cm}
\includegraphics{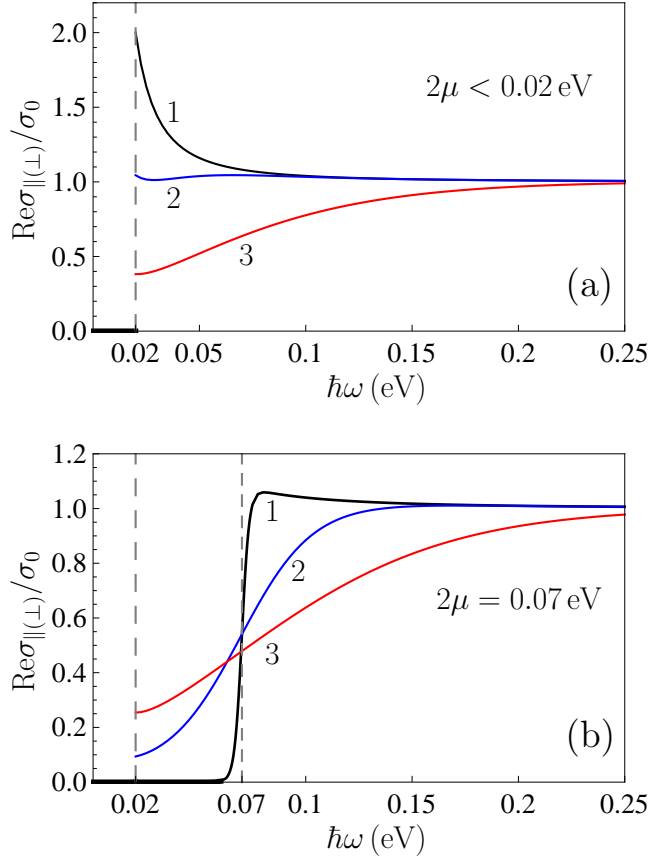}
}
\vspace*{-16cm}
\caption{\label{fg1}
The real part of the conductivity of graphene with the mass gap
$\Delta=0.02\,$eV normalized to the universal conductivity is
shown as a function of frequency at $T=10\,$K, 100\,K, and 300\,K
(lines 1, 2, and 3, respectively) for (a) any value of the chemical
potential satisfying a condition $2\mu< 0.02\,\mbox{eV}<\Delta$
and (b) $2\mu=0.07\,\mbox{eV}>\Delta$.
}
\end{figure}
\begin{figure}[b]
\vspace*{-0cm}
\centerline{\hspace*{2.5cm}
\includegraphics{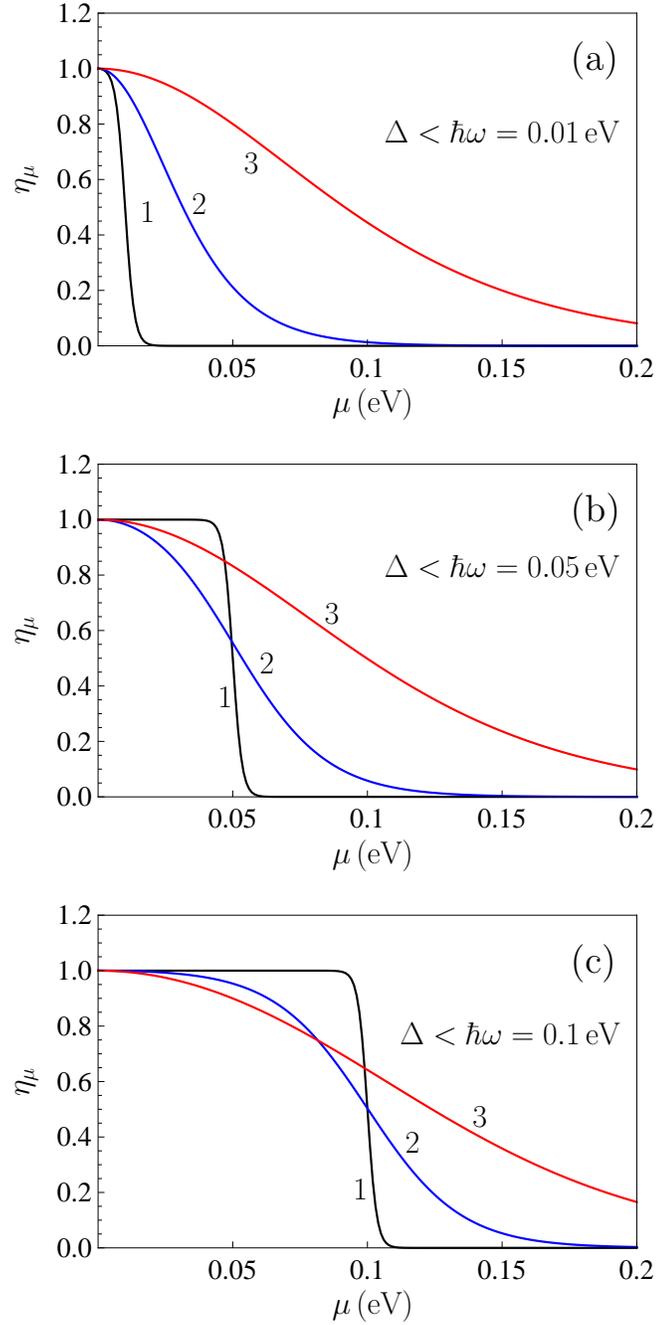}
}
\vspace*{-10cm}
\caption{\label{fg2}
The real part of the conductivity of graphene normalized to its
value at zero chemical potential is shown as a function of the chemical
potential at $T=10\,$K, 100\,K, and 300\,K
(lines 1, 2, and 3, respectively) for (a) $\ho=0.01\,$eV,
(b) $\ho=0.05\,$eV, and (c) $\ho=0.1\,$eV.
}
\end{figure}
\begin{figure}[b]
\vspace*{-0cm}
\centerline{\hspace*{2.5cm}
\includegraphics{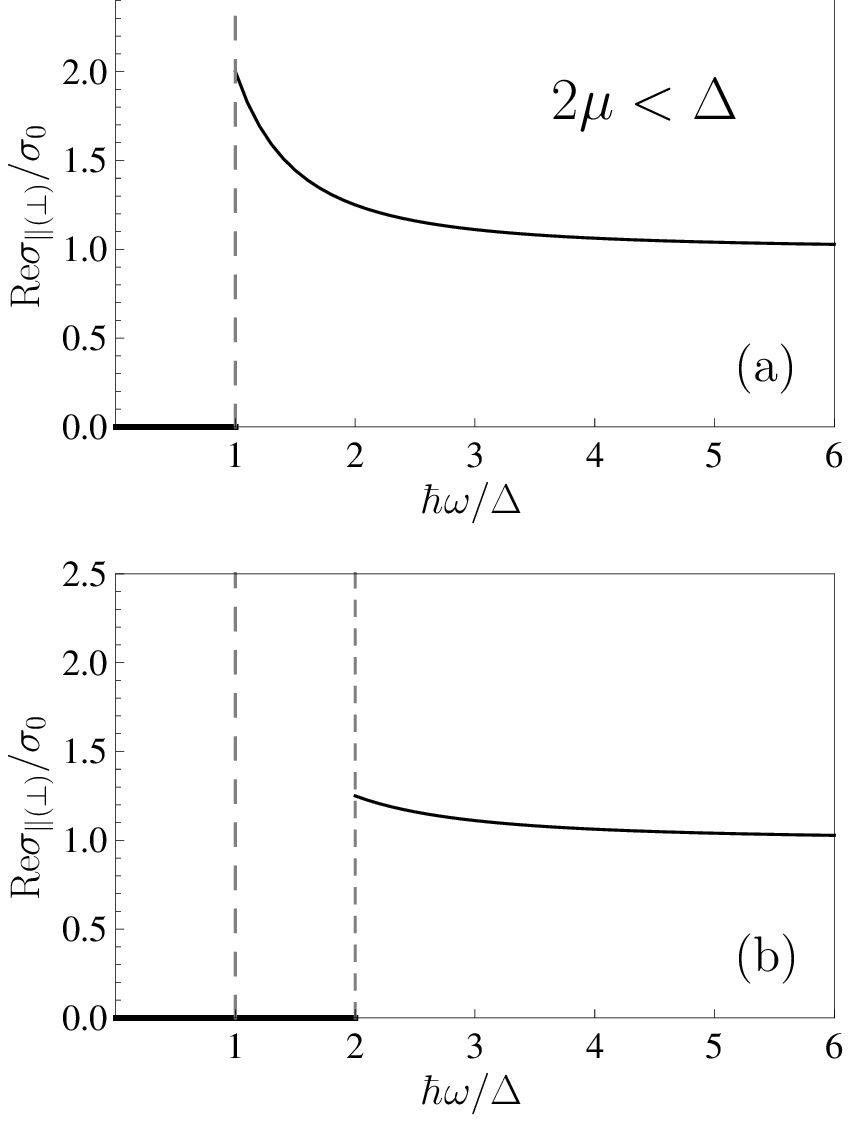}
}
\vspace*{-16cm}
\caption{\label{fg3}
The real part of the conductivity of graphene
 normalized to the universal conductivity is
shown as a function of $\ho/\Delta$ at zero temperature for (a) any
value of the chemical potential satisfying a condition $2\mu<\Delta$ and
(b) $\mu=\Delta$.
}
\end{figure}
\begin{figure}[b]
\vspace*{-8cm}
\centerline{\hspace*{2.5cm}
\includegraphics{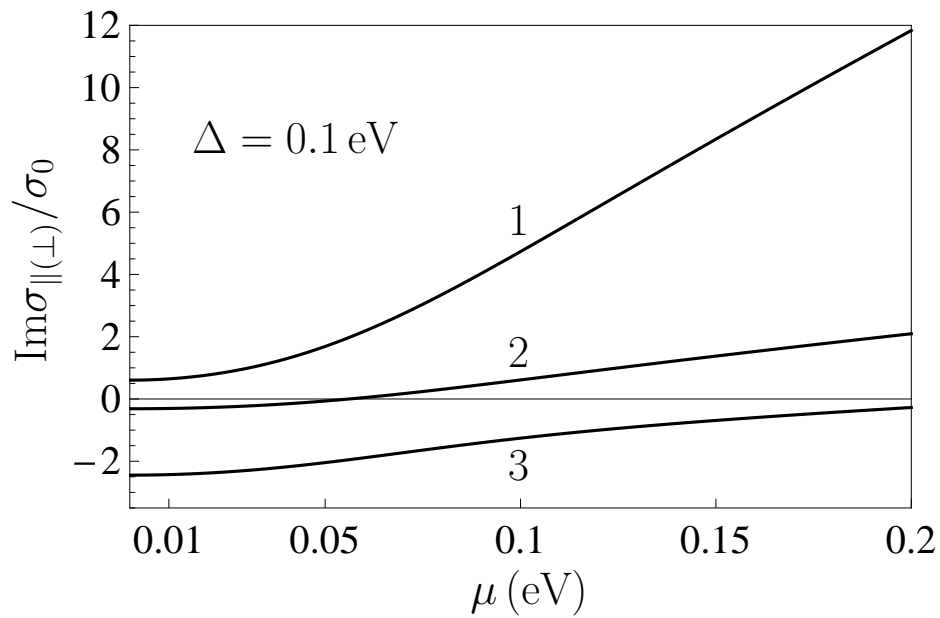}
}
\vspace*{-10cm}
\caption{\label{fg4}
The imaginary part of the conductivity of graphene normalized to
the universal conductivity  is shown as a function of the chemical
potential at $T=300\,$K for three values of frequency
 $\ho=0.01$, 0.05, and 0.099\,eV
(lines 1, 2, and 3, respectively).
}
\end{figure}
\begin{figure}[b]
\vspace*{-0cm}
\centerline{\hspace*{2.5cm}
\includegraphics{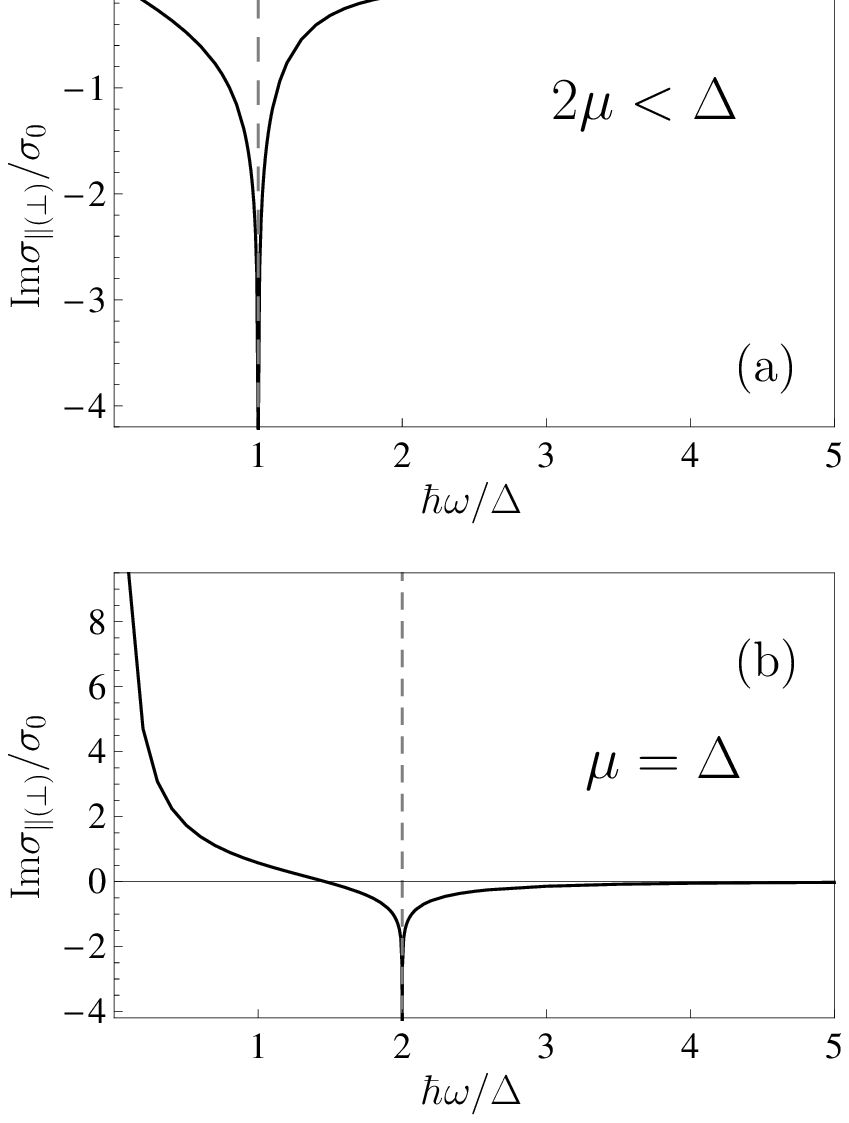}
}
\vspace*{-16cm}
\caption{\label{fg5}
The imaginary part of the conductivity of graphene normalized to
the universal conductivity  is shown as a function of
$\ho/\Delta$ at zero temperature for (a) any
value of the chemical potential satisfying a condition $2\mu<\Delta$ and
(b) $\mu=\Delta$.
}
\end{figure}
\end{document}